\documentclass[pdflatex,sn-mathphys-num]{sn-jnl}

\usepackage{graphicx}%
\usepackage{xcolor}%
\usepackage{colortbl}%
\usepackage{multirow}%
\usepackage[table]{xcolor} 
\usepackage{booktabs}
\usepackage{amsmath,amssymb,amsfonts}%
\usepackage{amsthm}%
\usepackage{mathrsfs}%
\usepackage{pifont}
\usepackage[title]{appendix}%
\usepackage{xcolor}%
\usepackage{tikz}
\usepackage{textcomp}%
\usepackage{manyfoot}%
\usepackage{booktabs}%
\usepackage{algorithm}%
\usepackage{algorithmicx}%
\usepackage{algpseudocode}%
\usepackage{listings}%

\theoremstyle{thmstyleone}%
\theoremstyle{thmstyletwo}%

\theoremstyle{thmstylethree}%

\begin{document}

\title[Article Title]{UNet with Self-Adaptive Mamba-Like Attention and Causal-Resonance Learning for Medical Image Segmentation}


\author[1,2]{\fnm{Saqib} \sur{Qamar}}

\author*[3]{\fnm{Mohd} \sur{Fazil}}\email{mfazil@imamu.edu.sa}
\author[4]{\fnm{Parvez} \sur{Ahmad}}
\author[3]{\fnm{Shakir} \sur{Khan}}
\author[5]{\fnm{Abu Taha} \sur{Zamani}}

\affil[1]{\orgdiv{Faculty of Computing and IT (FCIT)}, \orgname{Sohar University}, \orgaddress{\city{Sohar}, \postcode{311}, \country{Oman}}}

\affil[2]{\orgname{Department of Intelligent Systems}, \orgaddress{\street{KTH Royal Institute of Technology}, \postcode{10044}, \state{Stockholm}, \country{Sweden}}}

\affil[3]{\orgdiv{College of Computer and Information Sciences}, \orgaddress{\street{Imam Mohammad Ibn Saud Islamic University (IMSIU), Riyadh}, \country{Kingdom of Saudi Arabia}}}

\affil[4]{\orgdiv{Auckland Bioengineering Institute}, \orgname{University of Auckland}, \orgaddress{\street{Auckland}, \city{P. O. Box 1010}, \country{New Zealand}}}

\affil[5]{\orgdiv{Department of Computer Science, Faculty of Science}, \orgaddress{\street{Northern Border University, Arar 73213}, \country{Kingdom of Saudi Arabia}}}




\abstract{Medical image segmentation plays an important role in various clinical applications, but existing deep learning models face trade-offs between efficiency and accuracy. Convolutional Neural Networks (CNNs) capture local details well but miss global context, whereas Transformers handle global context but at high computational cost. Recently, State Space Sequence Models (SSMs) have recently shown potential for capturing long-range dependencies with linear complexity, but their direct use in medical image segmentation remains limited due to incompatibility with image structures and autoregressive assumptions. To overcome these challenges, we propose SAMA-UNet, a novel U-shaped architecture that introduces two key innovations. First, the Self-Adaptive Mamba-like Aggregated Attention (SAMA) block adaptively integrates local and global features through dynamic attention weighting, enabling efficient representation of complex anatomical patterns. Second, the Causal-Resonance Multi-Scale Module (CR-MSM) improves encoder–decoder interactions by adjusting feature resolution and causal dependencies across scales, enhancing semantic alignment between low- and high-level features. Extensive experiments on MRI, CT, and endoscopy datasets demonstrate that SAMA-UNet consistently outperforms CNN-, Transformer-, and Mamba-based methods. For example, it achieves 85.38\% DSC and  87.82\% NSD on BTCV, 92.16\% and 96.54\% on ACDC, 67.14\% and 68.70\% on EndoVis17, and 84.06\% and 88.47\% on ATLAS23, establishing new benchmarks across modalities. These results confirm the effectiveness of SAMA-UNet in combining efficiency with accuracy, making it a promising solution for real-world clinical segmentation tasks. The source code is available at \href{https://github.com/sqbqamar/SAMA-UNet}{GitHub}.}




\maketitle

\section{Introduction}\label{sec1}

Segmentation is a fundamental step in medical image analysis that enables the division of images into meaningful regions of interest. This process plays a crucial role in a wide range of medical applications, such as disease diagnosis, cancer microenvironment quantification, treatment planning, and disease progression tracking \cite{1}. Advanced methods like deep learning help researchers to examine different types of medical images to create accurate segmentation maps that highlight specific organs or areas of disease in diagnosis. Convolutional Neural Networks (CNNs) and Transformers are advanced deep learning approaches in the domain of medical image segmentation. CNNs use shared weights to excel in capturing translation invariance and local feature extraction. UNet \cite{2} is a prominent architecture that incorporates skip connections between the encoder and decoder that enable the integration of low-level and high-level details, which enhances its efficiency in extracting hierarchical image features for high-resolution medical images. However, CNNs rely on local convolutional kernels and struggle with capturing long-range dependencies, which limits their ability to handle global context in medical image segmentation.

Transformers, originally developed for natural language processing (NLP), have been successfully adapted to computer vision tasks, including medical image segmentation. Vision Transformers (ViT) \cite{3} and Swin Transformers \cite{4} are excellent at capturing global dependencies because they interpret images as sequences of tokens. Hybrid models such as TransUNet \cite{5} and SwinUNETR \cite{6} combine the local feature extraction of CNNs with the global modeling strengths of Transformers that lead to improved segmentation performance. However, Transformers produce computational complexity with respect to the sequence length, which becomes prohibitive when dealing with high-resolution medical images that require processing long sequences. State Space Sequence Models (SSMs) \cite{7, 8} have gained significant attention for their ability to model long-range dependencies with linear computational complexity. These models have achieved success in NLP tasks and have been extended to computer vision, where approaches like Visual State Space (VSS) \cite{9}, bi-directional SSM \cite{10}, and omnidirectional SSM \cite{11} have demonstrated promising results. Studies have highlighted that the Mamba module \cite{13}, a macro design within SSMs, significantly contributes to feature extraction in image-based tasks. However, while Mamba has achieved success in medical image segmentation, it also faces inherent difficulties. Specifically, Yu and Wang \cite{12} has presented Mamba's inability to incorporate causal relationships between image tokens limits its application in autoregressive tasks such as segmentation. Yu and Wang \cite{12} highlighted that one of the key limitations of Mamba is its inability to incorporate causal relationships between image tokens, which restricts its applicability to autoregressive tasks such as segmentation. In medical image segmentation the causal relationship can be understood in a spatial and hierarchical way where fine details at higher resolutions naturally come first and support the formation of coarser semantic features. Preserving this order is important to keep encoder and decoder features consistent. Recent studies have emphasized the importance of advanced feature fusion strategies to improve segmentation accuracy in complex medical imaging scenarios. Ju et al. \cite{jul} proposed CLDSINet, a collaborative learning framework that jointly models static and dynamic information using multi-branch feature extraction to enhance feature representation. This development motivates our design of the CR-MSM module, which explicitly preserves hierarchical dependencies across scales while maintaining computational efficiency. Similarly, Multi-View Convolutional Networks (MVCNs) \cite{mvcns} have demonstrated the benefit of combining multiple directional or perspective views for improved feature learning, but their reliance on parallel CNN branches makes them computationally expensive for high-resolution segmentation tasks.

To address these limitations, we introduce SAMA-UNet, a novel architecture designed to integrate Mamba-inspired innovations into transformer-based models. SAMA-UNet presents the SAMA block that combines contextual self-attention with dynamic feature modulation to enhance both local and global feature representations. This approach improves segmentation accuracy and reduces the computational complexity traditionally associated with transformers. Furthermore, we propose the CR-MSM, which enhances the integration of multi-scale features by introducing causal resonance learning where fine-grained encoder features resonate with semantically richer decoder features through a controlled multi-scale fusion process. This ensures that hierarchical dependencies across scales are not disrupted, thereby improving feature alignment and overall segmentation accuracy. This module encourages the inherent multi-scale feature generation in U-shaped networks to preserve causal dependencies while optimizing feature representation across scales. In addition, we address the quadratic complexity of transformers by incorporating a softmax attention mechanism inspired by the human foveal vision system. In human vision, the fovea prioritizes high-resolution detail in the center of focus while processing peripheral information more broadly and efficiently \cite{kolb}. Inspired by this biological principle, our pixel-focused attention dynamically allocates attention weights to give more importance to local fine details while still capturing global context and reducing unnecessary computations. We further employ differential attention to suppress irrelevant noise tokens and flash attention to accelerate efficiency on modern hardware. These innovations collectively allow SAMA-UNet to achieve high-performance segmentation while maintaining computational efficiency. In this paper, we propose the following contributions:

\begin{itemize}
    \item We present the SAMA Block, inspired by human foveal vision. Splitting channels into local-neighborhood and condensed global context branches enables linear computational complexity, while differential attention suppresses irrelevant tokens. By integrating Mamba’s macro-structure, SAMA maximizes channel-wise information use to yield richer feature representations.
    \item We suggest the CR-MSM to enhance feature fusion in U-shaped networks and improve the consistency of multi-scale information flow. By adjusting the SS2D token-scanning strategy to preserve causal order across scales, CR-MSM aligns with the autoregressive assumption. This design facilitates the complementary information among the various scales and effectively narrows the semantic gap between encoder and decoder features.
    \item We present a robust SAMA-UNet architecture for medical image segmentation that demonstrates superior performance across MRI, CT, and endoscopy datasets compared to existing CNN-based, transformer-based, and Mamba-based models.
\end{itemize}

Experimental results on four diverse medical imaging datasets demonstrate that SAMA-UNet significantly outperforms existing methods, offering a more efficient and accurate approach to medical image segmentation. The remain paper is organized as follows: In Sec.\ref{sec2}, we review related works, including mamba-based methods and transformer architectures for medical image segmentation. Sec.\ref{sec3} describes the details of our proposed method. Sec.\ref{sec4} presents the experimental setup, quantitative and qualitative results, and ablation studies. Finally, Sec.\ref{sec5} provides a comprehensive discussion of the paper. Finally, Sec.\ref{sec6} describes the conclusion of the paper.

\section{Related Work}\label{sec2}

\subsection{Transformers-based methods}

Transformers have become a leading architecture in natural language processing (NLP), computer vision, and multimodal tasks due to their ability to capture long-range dependencies and represent complex features. However, the quadratic computational complexity of traditional transformer models when processing high-resolution images remains a significant challenge. Researchers have proposed several methods to mitigate this issue while maintaining performance \cite{36}. Sparse attention techniques aim to reduce computational complexity by shortening token sequences. For example, shifted window-based attention \cite{4, 14} computes attention maps within smaller windows and extends global attention coverage through overlapping sliding windows. Similarly, Dilated Attention \cite{15} employs dilated rates to capture short- and long-range dependencies with limited sampling points. Deformable attention \cite{16, 17, 18} dynamically adjusts sampling positions based on learnt offsets to allow more flexible attention allocation tailored to specific input features. Furthermore, pixel-focused attention simulates how human vision works, allowing each feature point to see both local and global details. In addition, linear attention replaces the softmax operation in standard attention with a non-negative mapping function to reduce the complexity from quadratic to linear time. However, such an approach often leads to performance degradation. Building on this, focused linear attention \cite{20} improves the representations of the attention matrix by designing efficient mapping functions. Additionally, mamba-like linear attention \cite{13} combines SSM and linear attention in a larger design inspired by Mamba, which further boosts performance. Finally, flash attention \cite{21} makes the data transfer and computation processes for softmax attention on GPUs faster and more efficient, while still maintaining good performance. Flash attention \cite{21} accelerates GPU computation for softmax attention, improving efficiency while maintaining accuracy. 

Transformers have become a leading architecture in natural language processing (NLP), computer vision, and multimodal tasks due to their ability to capture long-range dependencies and represent complex features. However, the quadratic computational complexity of traditional transformer models when processing high-resolution images remains a significant challenge. Researchers have proposed several methods to mitigate this issue while maintaining performance \cite{36}. Sparse attention techniques aim to reduce computational complexity by shortening token sequences. For example, shifted window-based attention \cite{4,14} computes attention maps within smaller windows and extends global attention coverage through overlapping sliding windows. Similarly, Dilated Attention \cite{15} employs dilated rates to capture short- and long-range dependencies with limited sampling points. Deformable attention \cite{16,17,18} dynamically adjusts sampling positions based on learnt offsets to allow more flexible attention allocation tailored to specific input features. Pixel-focused attention simulates how human vision works, allowing each feature point to see both local and global details. In addition, linear attention replaces the softmax operation in standard attention with a non-negative mapping function to reduce the complexity from quadratic to linear time, although often at the expense of performance. Building on this, focused linear attention \cite{20} improves the representation of the attention matrix by designing efficient mapping functions. Mamba-like linear attention \cite{13} combines SSM and linear attention in a larger design inspired by Mamba, further boosting performance. Flash attention \cite{21} accelerates GPU computation for softmax attention, improving efficiency while maintaining accuracy. In addition to these attention optimizations, recent works have also emphasized adaptive feature fusion to improve efficiency. For instance, AFBNet \cite{yin2024afbnet} proposed a lightweight adaptive feature fusion module that balances local and global context, while DeepU-Net \cite{zhou2025deepu} employed a dual-branch design for multiscale fusion in remote sensing tasks. Such approaches highlight the trend toward efficient integration of features across scales, which motivates the design of our CR-MSM module.

These advancements have inspired us to develop SAMA-UNet, which combines Mamba-like macro-architectural design with pixel-focused attention and hardware-accelerated differential attention. This enables self-attention to achieve more efficient computational complexity while preserving high representation capacity, particularly for high-resolution medical image segmentation.

\subsection{Hybrid CNN–Transformer methods}

CNNs have been the backbone of medical image segmentation because of their strength in capturing local spatial features. Classical models such as U-Net \cite{2} and its variants U-Net++ \cite{qamar2021dense}, Attention U-Net \cite{saqib1}, and Dense U-Net \cite{40} achieve strong performance through hierarchical feature extraction and skip connections that improve fine-grained localization. However, CNNs are limited by their local receptive fields and struggle to capture long-range dependencies in high-resolution images. 

To address this, hybrid CNN–Transformer models have emerged, combining CNNs’ local detail extraction with Transformers’ global context modeling. TransUNet \cite{5} embedded a Transformer encoder into U-Net, while SwinUNETR \cite{6} used hierarchical Swin Transformer blocks for multi-scale representation. More recently, methods such as EFF-ResNet-ViT \cite{effersvit}, ResLNet\cite{wang2022reslnet}, and DCSSGA-UNet \cite{hus} have explored efficient CNN–Transformer hybrids that integrate attention and fusion mechanisms to selectively enhance critical features while reducing redundancy. These hybrids achieve a balance between local and global learning but often introduce high computational cost. Our proposed SAMA-UNet builds on this line of work by embedding a self-adaptive Mamba-like attention mechanism into a CNN–Transformer-inspired design. Unlike existing hybrids, SAMA-UNet explicitly models causal dependencies across scales with the CR-MSM module, enabling more efficient feature alignment while maintaining computational efficiency.

\subsection{Mamba-based methods}
State Space Sequence Models \cite{7, 8} have shown promising results in modelling long-range dependencies with linear computational complexity in NLP and computer vision. The Mamba module, a specific SSM-based method, introduces a selective mechanism that enhances context-based reasoning by enabling input-dependent interactions along sequences via linear transformations. Mamba’s macro-architectural design has been applied in hybrid CNN-SSM structures to improve image processing tasks, particularly in medical image segmentation \cite{37}. Several extensions of Mamba have been proposed to improve segmentation performance. For example, U-Mamba \cite{22} integrates Mamba blocks after convolutional layers into the U-Net encoder, outperforming traditional CNN and transformer-based models. Swin-UMamba \cite{23} and VM-UNet \cite{24} replace convolutional layers in U-Net with components inspired by Visual State Space, improving the representation of local and global features. VM-UNet-V2 \cite{25} refines skip connections using the convolutional block attention mechanism and introduces a Semantics and Detail Infusion module to strengthen the interaction between low- and high-level features. At the same time, LKM-UNet \cite{26} uses patch-based methods along with Mamba modules to extract features at both local and global levels. Other recent CNN–attention hybrid designs have also targeted the semantic gap problem. For example, DCSSGA-UNet \cite{hus} combines DenseNet201 with channel–spatial attention (CSA) and semantic guidance attention (SGA) to selectively enhance critical features while reducing redundancy between encoder and decoder pathways. Additionally, MSVM-UNet \cite{27} and MSM-UNet \cite{38} adds multi-scale convolutions into Visual State Space blocks to effectively capture 2D features at different scales. 

Although these models successfully integrate Mamba into medical image segmentation, they do not fully exploit Mamba’s advantages in analyzing causal sequential data. Our work builds upon the strengths of Mamba by focusing on the latent causal relationships between multi-scale features in U-Net. By embedding causal resonance learning in the encoder-decoder structure, we enhance feature representation and bridge the semantic gap between the encoder and decoder output, leading to better performance in high-resolution medical image segmentation.

Our SAMA-UNet model makes these existing hybrid models better by adding SAMA, which adjusts features dynamically and uses causal resonance learning, resulting in improved segmentation accuracy and less computational effort. By combining these innovations, SAMA-UNet addresses the inherent challenges of traditional CNNs, Transformers, and Mamba-based methods, offering a more efficient and robust solution for high-resolution medical image segmentation tasks.

\section{Methodology}\label{sec3}

\begin{figure}[] 
    \centering 
    \includegraphics[width=\textwidth]{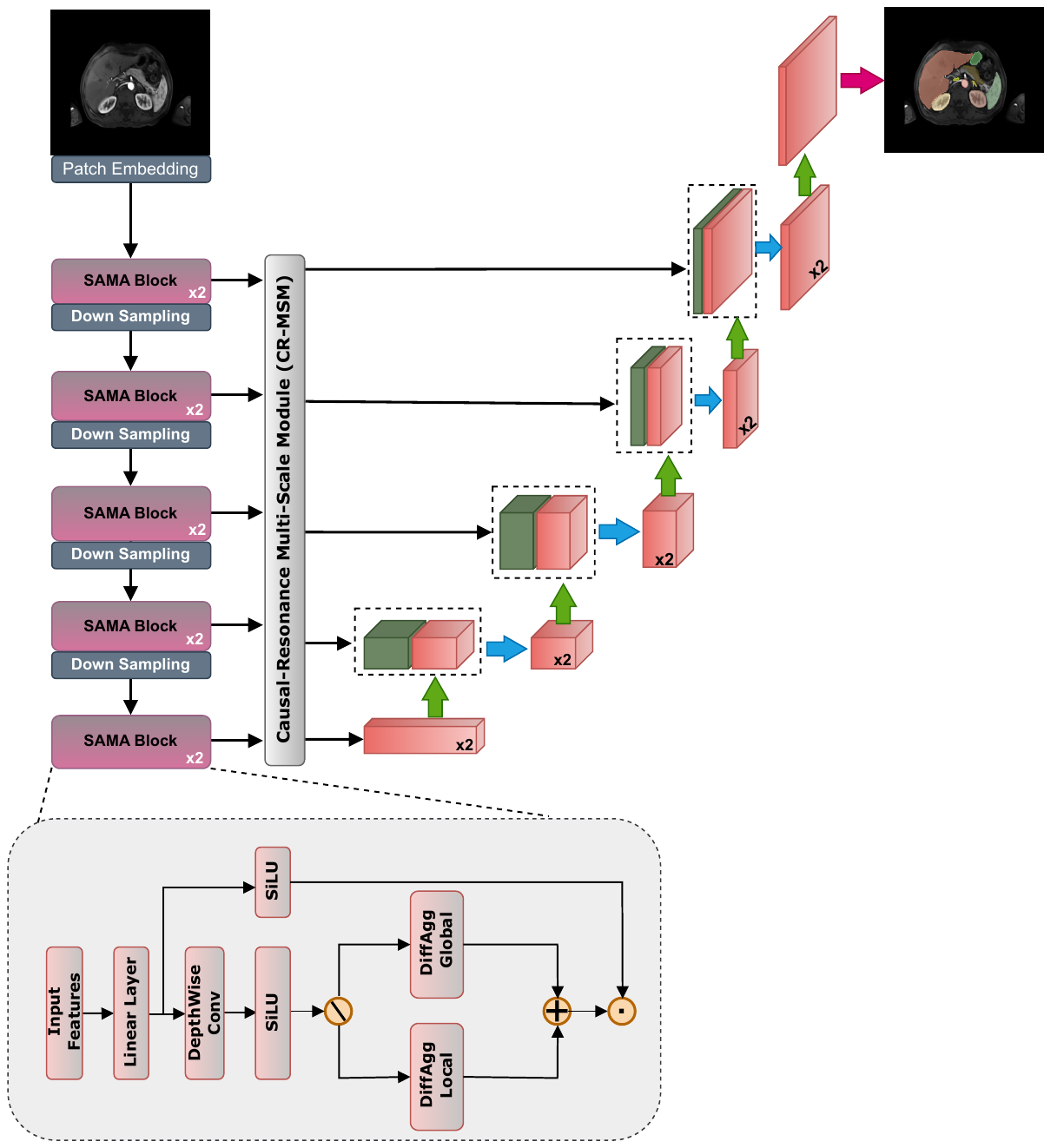} 
    \caption{Overview of the SAMA-UNet architecture. SAMA is used in the encoder, residual convolution blocks are used in the decoder, and a CR-SMM with implicit causality is applied in the skip connections. The SAMA block includes a token mixer sub-block with a modified pixel-focused attention mechanism and a Feed-Forward Neural Network (FFN) sub-block.}
    \label{fig1} 
\end{figure}

The proposed SAMA-UNet architecture is designed to enhance the efficiency and accuracy of medical image segmentation, as illustrated in Figure \ref{fig1}. Our model incorporates several key innovations to address the computational inefficiencies and challenges posed by medical data, allowing more precise segmentation of complex anatomical structures. Specifically, SAMA-UNet integrates the following components: (1) SAMA block replaces the SSM in the Mamba module with a softmax attention mechanism featuring linear computational complexity, which effectively eliminates the need for the autoregressive assumption. (2) CR-MSM takes the advantages of the implicit causal relationships seen in the continuous multi-scale features of U-shaped networks and introduces a novel scan-expand method to enhance the 2D-Selective-Scan (SS2D) operation. This design better aligns the continuous long-sequence modeling capabilities of SSMs with the requirements of medical image segmentation tasks.

\subsection{Backbone architecture}
The architecture of SAMA-UNet follows a U-shaped design for efficient medical image segmentation. The patch embedding module first processes the input image, transforming it into feature embeddings. It uses overlapping patches to break down the image by reducing spatial dimensions and increasing channel dimensions through convolutional operations. This procedure ensures the preservation of important spatial information while embedding the image into higher-dimensional features. The embedded features are then passed through a series of SAMA blocks in the encoder, where they undergo dynamic attention processing. The SAMA block reduces the computational complexity of the traditional attention mechanism while also enhancing the model’s ability to capture both local and global dependencies. These features are further refined through a modified skip connection that includes the causal-resonance multi-scale module to better align the information between the encoder and decoder. The multiscale optimized features are progressively fused in the decoder via convolutions and gradually restored to the original image resolution through transpose convolutions. Finally, the segmentation results are generated by the decoding heads at different scales.

\subsection{SAMA}

\begin{figure}[h]
    \centering
    \includegraphics[width=1.1\textwidth]{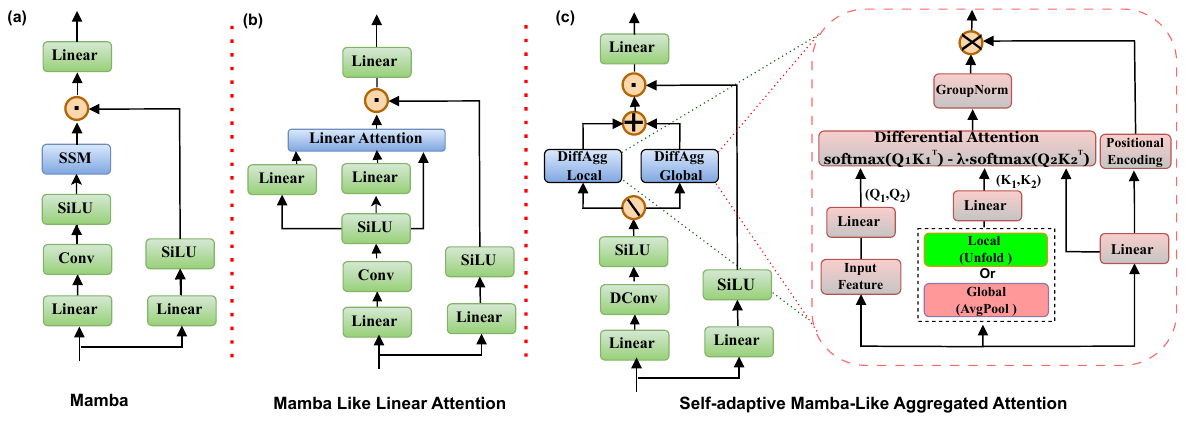}
    \caption{Illustration of the structure designs of Mamba, Mamba-Like Linear Attention, and our SAMA block.}
    \label{fig2}
\end{figure}

The SAMA block is a key innovation in SAMA-UNet, designed to enhance transformer efficiency while maintaining high representation capacity. The block is built upon the foundation of Mamba-like attention mechanisms, combining the advantages of contextual self-attention and dynamic weight modulation. Figure \ref{fig2}(c) illustrates the proposed SAMA block. Figure \ref{fig2} also highlights the differences between our module and other modules based on Mamba. As shown in Figure \ref{fig2}(a), Mamba uses SSMs and Gated Attention \cite{28} to change its design, while MLLA combines Linear Attention with Mamba’s structure, as seen in Figure \ref{fig2}(b). For clarity and simplicity in the illustration, the multiple attention heads used are not depicted in the figure.

Drawing inspiration from \cite{12, 13}, the proposed SAMA integrates multiple Mamba-like design elements into the macro-structure of the attention mechanism. These include a SiLU activation layer, a depthwise convolution layer (DepthConv) and a linear layer preceding the attention layer, as well as bypass activation and weighting strategies. These designs enhance the model’s ability to capture both positional and channel information to improve the encoder performance. The operation within the SAMA block involves two branches: the local branch and the global branch. The local branch focuses on local feature extraction, while the global branch aggregates global context. The local and global features are processed in parallel using Aggregated Attention and then concatenated for further processing. This approach reduces computational cost while ensuring the effective capture of both local and global features.

The procedure of the SAMA blocks can be described as Equation 1. $X_l$  and $X_g$  represent the inputs of the local and global branches, respectively. Split and Concat refer to the split and concatenate operations along the channel dimension.

\begin{align}
X &= \text{SiLU}(\text{DepthConv}(\text{Linear}(I))) \notag \\
X_l, X_g &= \text{Split}(X) \notag \\
X_{res} &= \text{SiLU}(\text{Linear}(I)) \tag{1} \\
O &= \text{Linear}(\text{Concat}(\text{DiffAgg}(X_l), \text{DiffAgg}(X_g)) \cdot X_{act\_res}) \notag
\end{align}

In addition, we introduced several modifications to the pixel-focused attention method. Firstly, we split the feature map along the channel dimension into two parts: one for local attention and another for global attention. These components are processed separately and then concatenated. This design addresses two key issues: (1) directly stacking both local and global attention may introduce inconsistent biases for features spanning local and global ranges due to down-sampling; (2) splitting the feature map reduces computational complexity. Second, we incorporated locally enhanced positional encoding (PE) into the softmax attention module. This modification provides positional information using convolution while removing the relative positional bias. Third, we eliminated certain learnable embedding parameters and other configurations to simplify the attention module.

Finally, inspired by the Differential Transformer \cite{35}, we introduced differential attention into the computation of both local and global attentions. Their findings indicate that some non-negligible attention scores are assigned to irrelevant contexts in the attention maps. As a result, the irrelevant context dilutes the weight distribution of relevant information, negatively impacting overall performance. By incorporating this structure, we applied differential denoising to our attention module, helping it cancel out attention noise and focus more effectively on relevant information.

\[
\begin{cases}
Q_{l,(i,j),1}, Q_{l,(i,j),2} &= \text{Split}(Q_{l,(i,j)}) \\
K_{l,\rho(i,j),1}, K_{l,\rho(i,j),2} &= \text{Split}(K_{l,\rho(i,j)})
\end{cases} \tag{2}
\]
\[
\begin{cases}
A_{l,(i,j),1} &= \text{Softmax}\left(\frac{Q_{l,(i,j),1}K_{l,\rho(i,j),1}^T}{\sqrt{d/2}}\right) \\
A_{l,(i,j),2} &= \text{Softmax}\left(\frac{Q_{l,(i,j),2}K_{l,\rho(i,j),2}^T}{\sqrt{d/2}}\right) \\
\text{DiffAttn}(Q_{l,(i,j)}, K_{l,\rho(i,j)}, V_t, \lambda) &= 
(A_{l,(i,j),1} - \lambda A_{l,(i,j),2})V_t
\end{cases} \tag{3}
\]

We describe the proposed Differential Aggregated Attention module as follows, using the local branch as an example: First, as shown in Equation 2, the $Q$ and $K$ tokens of the local branch are split along the channel dimension into $Q_1$, $Q_2$ and $K_1$, $K_2$, respectively. Then, Equation 3 shows how we create differential attention by taking the difference between two sets of Softmax Attention, using a learnable scalar $\lambda$, which starts at a value of $\lambda_{\text{init}}$. $d$ represents the channel dimension of features after the Split operation in Equation 1. Following the parameter settings in \cite{35}, $\lambda_{\text{init}}$ is set to a fixed value of 0.8. Finally, we apply a Group Normalisation (GN) operation to the differential attention, followed by a scalar dot product to generate the differential output. This output is then combined with the positional information generated by PE to produce the Differential Aggregated Attention (DiffAgg) output.

\subsection{CR-MSM}

The CR-MSM is designed to enhance multi-scale feature fusion in SAMA-UNet by explicitly preserving the natural order of information flow across scales. Traditional fusion strategies, such as simple concatenation and convolutional aggregation, often mix high- and low-level features without regard to their hierarchical relationship, which disrupts semantic consistency between the encoder and decoder. CR-MSM addresses this limitation through causal resonance learning, where information from fine-grained features resonates with semantically richer features at coarser scales, ensuring that feature fusion reinforces rather than distorts the natural progression from detail to abstraction. Unlike MVCNs, which require parallel CNN branches or external multi-view inputs to capture diverse perspectives, CR-MSM generates directional views internally from the same feature map and fuses them efficiently using State Space Modeling. This design avoids the heavy computational overhead of MVCNs while achieving more consistent feature alignment.

In U-shaped networks such as UNet, encoder features at higher resolutions naturally precede and support the formation of lower-resolution features. This hierarchical dependency forms a causal chain where local fine details act as the cause and global semantic features are the effect. CR-MSM explicitly models this causal dependency by aligning encoder and decoder features so that the structural flow of information is preserved. By doing so, the module narrows the semantic gap across scales and improves segmentation accuracy, especially for complex anatomical patterns. To achieve this, CR-MSM introduces directional multi-view transformations combined with SSM.  Each encoder feature map is transformed in four orientations which are original, transposed, flipped, and flipped-transposed to capture spatial dependencies from multiple perspectives. These directional views are then flattened in 1D sequences and individually processed through an SSM block, which employs structured linear recurrence to model long-range dependencies with linear computational complexity. This design allows CR-MSM to efficiently capture contextual relationships across orientations and avoiding the quadratic cost of self-attention while retaining strong representational power.

After SSM processing, the outputs of the four directional views are reconstructed to their original orientation and fused through causal resonance averaging. Unlike simple concatenation, which may merge features indiscriminately, causal resonance averaging respects the inherent order of feature generation and ensures that fine details and abstract features reinforce one another. Formally, for feature maps $F_1, F_2, \ldots, F_N$ at different scales, the transformed sequences $X_1, X_2, X_3, X_4$ are processed into outputs $Y_1, Y_2, Y_3, Y_4$ by the SSM, and the final fused feature map is given by:

\[
Y_{\text{final}} = \frac{1}{4} \left( Y_1 + Y_2 + Y_3 + Y_4 \right).
\]

This fused representation is then projected through a linear layer and passed to the decoder for segmentation. By enforcing directional consistency and resonance across scales, CR-MSM maintains the causal flow of information within the U-shaped architecture and produces multi-scale features that are semantically aligned, computationally efficient, and highly effective for medical image segmentation. The detailed procedure of the proposed CR-MSM is summarized in Algorithm~\ref{crmsm}. In which, $\{F_i\}_{i=1}^N$ are the encoder feature maps. Each $F_i$ is transformed into four directional views $\{X_j\}_{j=1}^4$, which are processed by SSM to produce $\{Y_j\}_{j=1}^4$. The fused output at scale $i$ is $\widetilde{F}_i$ through causal-resonance averaging, and $\{Z_i\}_{i=1}^N$ are the linearly projected skip features for the decoder.

\begin{algorithm}[htbp]
\caption{Causal-Resonance Multi-Scale Module (CR-MSM)}
\label{crmsm}
\begin{algorithmic}[1]
\Require Encoder feature maps $\{F_i\}_{i=1}^{N}$ 
\Ensure Fused skip features $\{Z_i\}_{i=1}^{N}$

\For{$i \gets 1$ to $N$}
  \State Generate four views: 
  $F_i^{\text{orig}}, F_i^{\text{T}}, F_i^{\text{flip}}, F_i^{\text{flipT}}$
  \State Flatten each view to sequences: $X_1,\ldots,X_4$
  \State Apply SSM: $Y_j \gets \mathrm{SSM}(X_j)$ for $j=1..4$
  \State Fuse by causal-resonance averaging:
  \[
    \widetilde{F}_i = \tfrac{1}{4}(Y_1 + Y_2 + Y_3 + Y_4)
  \]
  \State Linear projection for skip connection: 
  $Z_i \gets \mathrm{Linear}(\widetilde{F}_i)$
\EndFor
\State \Return $\{Z_i\}_{i=1}^{N}$
\end{algorithmic}
\end{algorithm}

\section{Experiments and Results}\label{sec4}
\subsection{Experimental settings}
\subsubsection{Datasets}
The BTCV dataset is part of the Synapse abdominal multi-organ segmentation task, introduced in the Multi-Atlas Labeling Beyond the Cranial Vault Workshop and Challenge \cite{29}. It comprises 30 abdominal CT scans, containing a total of 3779 axial contrast-enhanced CT slices. Each CT volume consists of 85 to 198 slices with a resolution of $512 \times 512$ pixels. Following the data split in previous work \cite{5}, 18 cases are used for training and 12 cases for testing. The default input image size is set to $224 \times 224$. This setting is a slight modification based on the nnUNet configuration \cite{34}. ACDC dataset is derived from the Automated Cardiac Diagnosis Challenge \cite{30} and contains 100 cardiac MRI scans, each segmented into three substructures: right ventricle (RV), myocardium (Myo), and left ventricle (LV). A random split of 80 MRI samples is used for training, while the remaining 20 samples are designated for testing. Consistent with the nnUNet settings, the default input image size is set to $256 \times 224$. EndoVis17 dataset is from the MICCAI 2017 EndoVis Challenge \cite{31}, which focuses on the segmentation of seven surgical instruments in endoscopic images. We adopt the official dataset split, where the training set consists of 1800 and 1200 image frames extracted from eight videos. The test set contains frames from two additional unseen videos. The input image size is set to $384 \times 640$, following the nnUNet configuration. ATLAS23 is a publicly available segmentation dataset~\cite{ 32} originates from the MICCAI 2023 ATLAS Challenge and includes 60 CE-MRI T1-weighted scan along with 60 liver and liver tumor segmentation masks. The annotations were generated by radiologists. The input image size is set to $320 \times 250$, in accordance with the nnUNet configuration.

\subsubsection{Evaluation metrics}
Following recommendations in Metrics Reloaded \cite{33} and U-Mamba \cite{22}, we used Dice Similarity Coefficient (DSC) and Normalized Surface Distance (NSD) for these semantic segmentation tasks. The DSC is a widely used metric for evaluating the overlap between predicted and ground truth segmentations. Mathematically, it is expressed as:
$$\text{DSC} = \frac{2|G \cap P|}{|G| + |P|}$$
where $P$ and $G$ represent the predicted and ground truth segmentation regions, respectively.
The NSD quantifies the discrepancy between the surfaces of the predicted and ground truth segmentations. It focuses on boundary alignment, making it particularly useful for assessing segmentation quality in regions with fine structural details. It calculates the percentage of boundary points of the predicted segmentation that lie within a predefined tolerance distance from the ground truth boundary, normalized to the total boundary points. Mathematically, it can be expressed as:

\[
\text{NSD} = \frac{|\{x \in S_P : \text{dist}(x, S_G) \leq \tau\}| + |\{y \in S_G : \text{dist}(y, S_P) \leq \tau\}|}{|S_P| + |S_G|}
\]

Here, \( S_P \) denotes the set of surface points from the predicted segmentation, and \( S_G \) denotes the set of surface points from the ground truth segmentation. The function \( \text{dist}(x, S) \) represents the shortest distance from point \( x \) to the surface \( S \). The parameter \( \tau \) is a predefined tolerance value, and \( |\cdot| \) denotes the cardinality, i.e., the number of points in the set.

\subsubsection{Implementation details}
All experiments were conducted using Python 3.10 and PyTorch 2.0.0 on an NVIDIA GeForce RTX 3080 GPU with 24 GB memory. The nnUNet framework serves as the backbone, with 500 training epochs, each consisting of 250 iterations. The preprocessing and data augmentation methods strictly follow the methods implemented in nnUNet for the respective datasets. For a fair comparison, we implemented SwinUNETR, U-Mamba, LKM-UNet, and MLLA-UNet into the nnUNet framework. All networks were trained from scratch using the AdamW optimizer, with the default initial learning rate set to $5e^{-4}$ or $1e^{-4}$, although certain models followed the learning rate settings specified in their released code. A one-cycle Cosine Annealing learning rate scheduler are applied to adjust the learning rate during training.

\subsection{Quantitative and Qualitative results}

Table I summarizes the quantitative 2D segmentation results across four medical image segmentation datasets. Compared to CNN-based, Transformer-based, and Mamba-based methods, our proposed SAMA model consistently achieves superior performance in both DSC and NSD metrics.

\begin{table}[h]
\centering
\caption{Quantitative Results Summary of Four Segmentation Tasks: Synapse Abdominal Multi-Organ Segmentation (BTCV), Automated Cardiac Diagnosis (ACDC), Instruments Segmentation In Endoscopy Images (ENDOVIS17), and ATLAS23 for Liver Tumor Segmentation. The Best Result In Each Dataset Is Highlighted In Bold. The Evaluation Metrics Are DSC (\%) And NSD (\%).}
\begin{tabular}{l cc cc cc cc}
\toprule
\rowcolor[gray]{0.9}
\textbf{Methods} & \multicolumn{2}{c}{\textbf{BTCV}} & \multicolumn{2}{c}{\textbf{ACDC}} & \multicolumn{2}{c}{\textbf{Endovis17}} & \multicolumn{2}{c}{\textbf{ATLAS23}} \\
\cmidrule(r){2-3} \cmidrule(r){4-5} \cmidrule(r){6-7} \cmidrule(r){8-9}
\rowcolor[gray]{0.9}
 & DSC & NSD & DSC & NSD & DSC & NSD & DSC & NSD \\
\midrule
nnUNet \cite{34} (2021)       & 84.93 & 87.26 & 91.85 & 96.34 & 62.36 & 63.83 & 80.22 & 85.46 \\
SwinUNETR \cite{6} (2022)     & 78.26 & 78.90 & 91.12 & 94.13 & 57.90 & 59.41 & 79.15 & 82.87 \\
LKM-UNet \cite{26} (2024)     & 84.45 & 87.50 & 89.71 & 92.10 & 62.39 & 63.91 & 81.09 & 85.49 \\
U-Mamba(Enc) \cite{22} (2024) & 84.46 & 86.67 & 88.99 & 91.63 & 65.45 & 66.86 & 82.79 & 86.79 \\
\rowcolor{gray!10}
\textbf{SAMA-UNet (ours)}     & \textbf{85.38} & \textbf{87.82} & \textbf{92.16} & \textbf{96.54} & \textbf{67.14} & \textbf{68.70} & \textbf{84.06} & \textbf{88.47} \\
\bottomrule
\end{tabular}
\end{table}

On the BTCV dataset, our model outperforms the second-best model U-Mamba by 0.92\% and 1.15\% in DSC and NSD, respectively. We follow U-Mamba’s evaluation standard, eight out of thirteen organ classes (Aorta, Gallbladder, Kidney (L), Kidney (R), Liver, Pancreas, Spleen, and Stomach) are included in the calculation. Mamba based methods demonstrate strong overall performance, with our model maintaining a lead of 0.92\% and 1.15\% over U-Mamba in DSC and NSD. For the ACDC dataset, where all methods achieve relatively high accuracy, our method still surpasses SwinUNETR by 0.32\% and 0.20\% in DSC and NSD, respectively. On the EndoVis17 dataset, where both CNN-based and Mamba-based methods perform well, our method achieves further improvements, surpassing U-Mamba by 1.69\% in DSC and 1.84\% in NSD. Finally, on the ATLAS23 dataset, our model secures improvement by 1.27\% and 1.68\% in DSC and NSD, respectively while compare with U-Mamba.

\begin{figure}[h]
    \centering
    \includegraphics[width=1\textwidth]{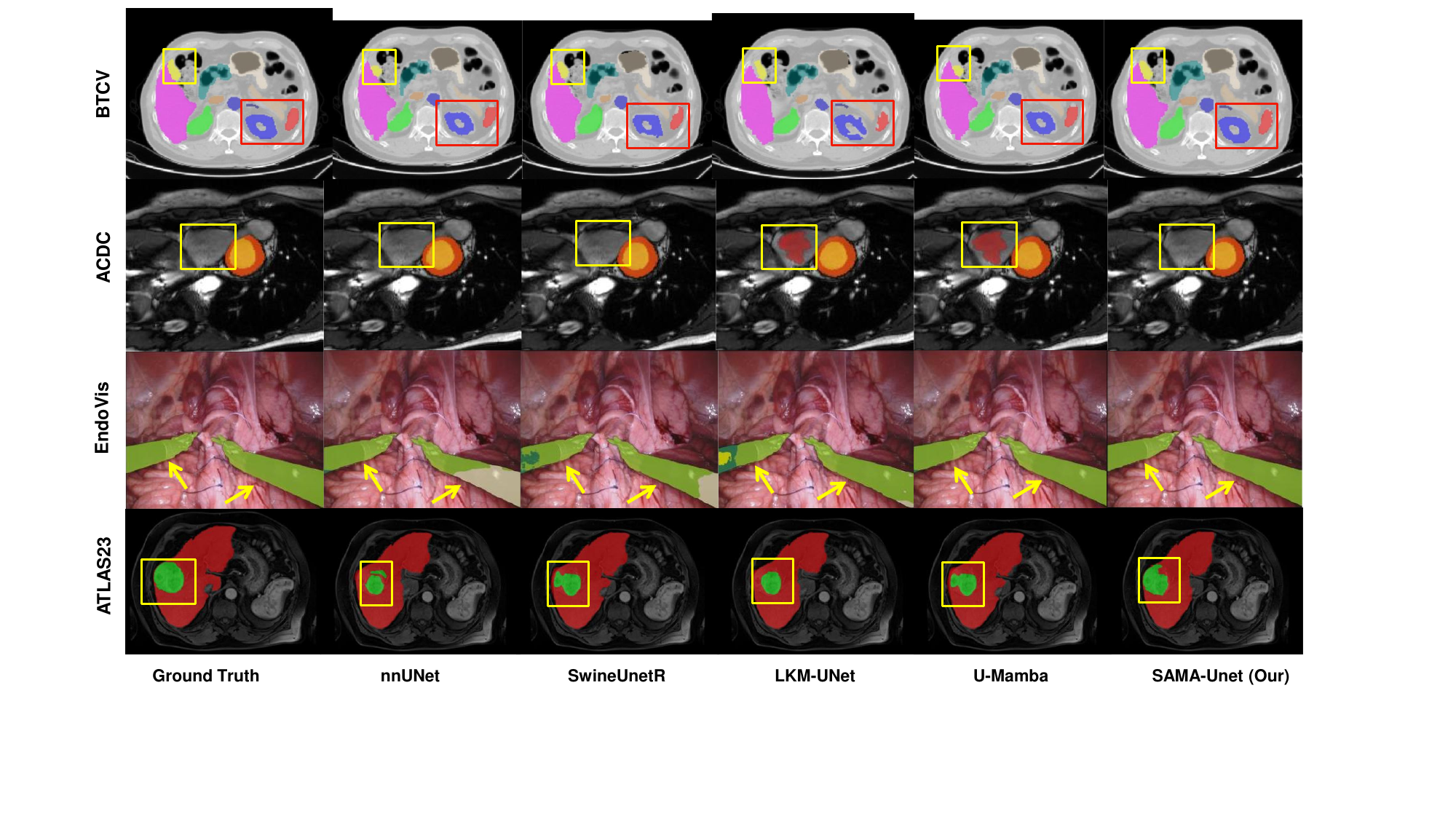}
    \caption{Visualization of segmentation examples Synapse multi-organ dataset (BTCV, 1st row), Automated Cardiac Diagnosis Challenge dataset (ACDC, 2nd), endoscopy images dataset (Endovis17, 3rd row), and liver tumor dataset (ATLAS23, 4th row). The result column of our proposed SAMA-UNet method is shown in the last column. SAMA-UNet is more robust to heterogeneous appearances and has fewer segmentation outliers.
}
    \label{fig3}
\end{figure}

These results collectively demonstrate that our model not only achieves strong generalization across diverse medical imaging modalities, including MRI, CT, and endoscopy, but also effectively enhances Transformer-based methods, resulting in significant performance improvements. Qualitative results from examples across four datasets are illustrated in Figure \ref{fig3}. In the first row, many methods, such as nnU-Net, SwinUNETR, U-Mamba, and LKMUNet, fail to accurately segment the contours of the organ’s regions due to their heterogeneous appearances. In the second row, methods like U-Mamba, and LKM-UNet produce segmentation errors in the right ventricle. In the third row, nnU-Net, SwinUNETR, and LKMUNet misclassify segments of the Prograsp Forceps, while U-Mamba produces misclassifications in the Bipolar Forceps and Ultrasound Probe. In the fourth row, several methods, including nnU-Net, SwinUNETR, U-Mamba, and LKMUNet, fail to accurately segment the contour of the liver tumor in the liver. In contrast, SAMA-UNet demonstrates clear advantages across these scenarios, highlighting its superior ability to capture both local details and global semantic information.

\subsection{Quantitative analysis of individual organ}   
We conducted a detailed quantitative analysis of individual categories on BTCV datasets, as shown in Table II. For individual organ segmentation on the BTCV dataset, our proposed SAMA-UNet achieves the highest DSC scores for several organs: 91.53\% for the Aorta, 68.95\% for the Gallbladder, 87.69\% for the Left Kidney, 86.37\% for the Right Kidney, 95.73\% for the Liver, 77.82\% for the Pancreas, 92.28\% for the Spleen, and 82.71\% for the Stomach. However, there is more potential for improvement in the segmentation of the Gallbladder and Pancreas.

\begin{table}[htbp]
\centering
\caption{Quantitative Results of Organ Segmentation in BTCV Dataset. The evaluation metrics are DSC (\%) and NSD (\%).}
\begin{tabular}{l cc cc cc cc cc}
\toprule
\rowcolor[gray]{0.9}
\textbf{Organs} & \multicolumn{2}{c}{\textbf{nnUNet}} & \multicolumn{2}{c}{\textbf{SwinUNETR}} & \multicolumn{2}{c}{\textbf{LKM-UNet}} & \multicolumn{2}{c}{\textbf{U-Mamba}} & \multicolumn{2}{c}{\textbf{SAMA-UNet (ours)}} \\
\cmidrule(r){2-3} \cmidrule(r){4-5} \cmidrule(r){6-7} \cmidrule(r){8-9} \cmidrule(r){10-11}
\rowcolor{gray!10}
 & DSC & NSD & DSC & NSD & DSC & NSD & DSC & NSD & DSC & NSD \\
\midrule
Aorta           & 91.61 & 95.12 & 89.30 & 93.11 & 91.03 & 94.64 & 91.47 & 95.43 & 91.53 & 95.09 \\
Gallbladder     & 69.93 & 69.75 & 61.44 & 59.21 & 65.32 & 69.49 & 68.76 & 69.42 & 68.95 & 69.58 \\
Left kidney     & 85.92 & 87.65 & 83.70 & 85.42 & 85.75 & 90.58 & 85.25 & 88.14 & 87.69 & 88.06 \\
Right kidney    & 83.85 & 86.74 & 83.92 & 89.28 & 93.99 & 91.13 & 94.24 & 89.05 & 86.37 & 88.36 \\
Liver           & 95.85 & 93.95 & 90.05 & 76.87 & 95.08 & 94.22 & 95.18 & 93.95 & 95.73 & 94.06 \\
Pancreas        & 77.80 & 87.18 & 66.51 & 67.25 & 68.13 & 79.89 & 68.10 & 79.90 & 77.82 & 89.17 \\
Spleen          & 90.31 & 89.59 & 85.70 & 81.20 & 89.70 & 93.92 & 89.08 & 92.39 & 92.28 & 93.46 \\
Stomach         & 84.20 & 88.62 & 66.65 & 74.80 & 84.33 & 83.90 & 83.60 & 85.13 & 82.71 & 84.76 \\
\midrule
\rowcolor{gray!10}
\textbf{Average} & \textbf{84.93} & \textbf{87.26} & \textbf{78.26} & \textbf{78.90} & \textbf{84.45} & \textbf{87.50} & \textbf{84.46} & \textbf{86.67} & \textbf{85.38} & \textbf{87.82} \\
\bottomrule
\end{tabular}
\label{btcv_organ}
\end{table}

\subsection{Ablation study} 
In this subsection, we conduct ablation studies on the BTCV dataset to verify the effectiveness of the key components in our method. First, we perform experiments to evaluate the effectiveness of the Mamba-Like Aggregated Attention module in our proposed method. Additionally, we conduct ablation experiments on certain configurations within the Multi-Scale Mamba Module with Implicit Causality.

\begin{figure}[h]
    \centering
    \includegraphics[width=0.8\textwidth]{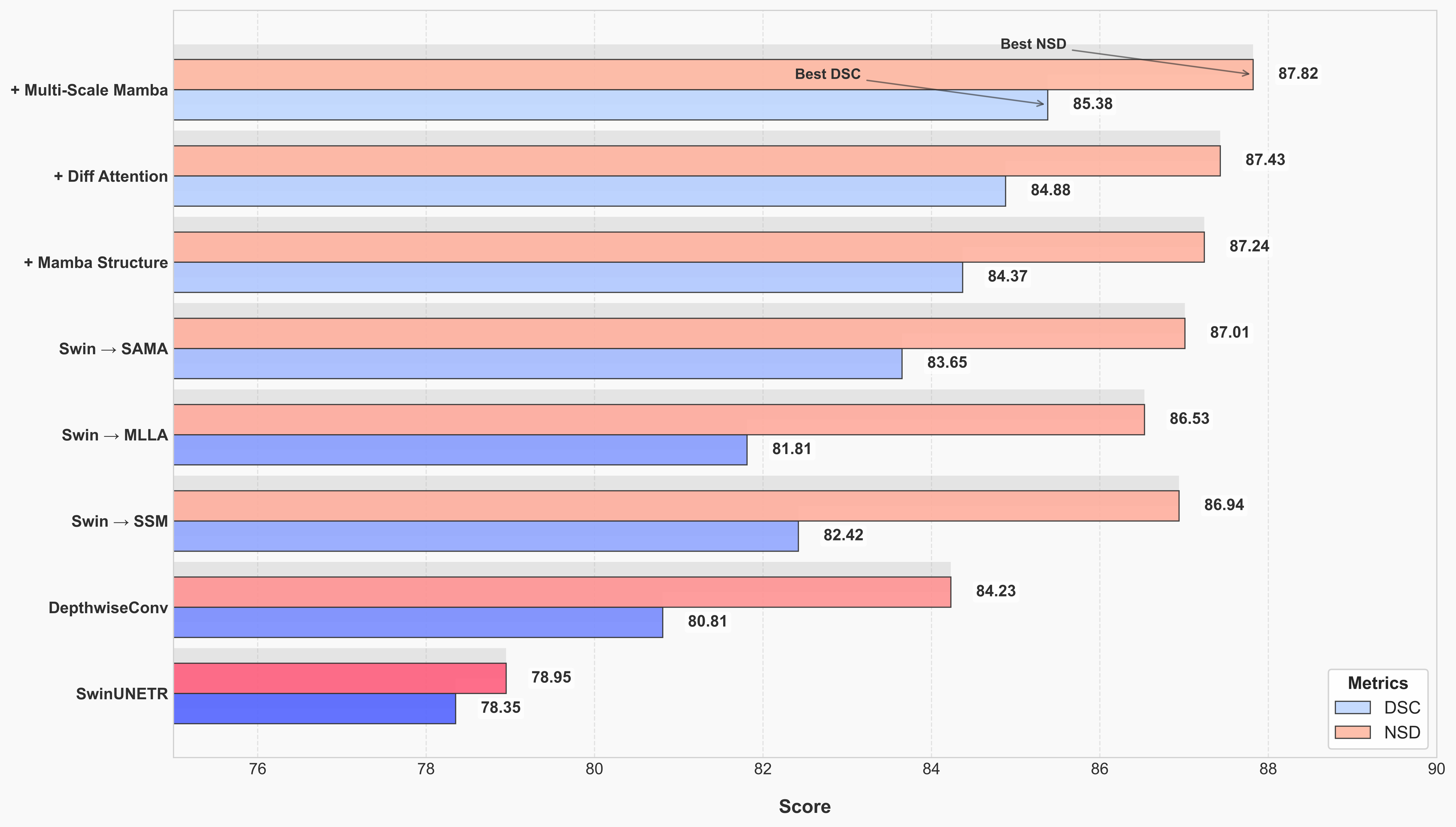}
    \caption{Illustration of performance variations observed through ablation studies, highlighting the impact of different architectural components on segmentation accuracy. Evaluations are based on DSC, NSD using the BTCV dataset.}
    \label{fig4}
\end{figure}

\begin{figure}[h]
    \centering
    \includegraphics[width=0.8\textwidth]{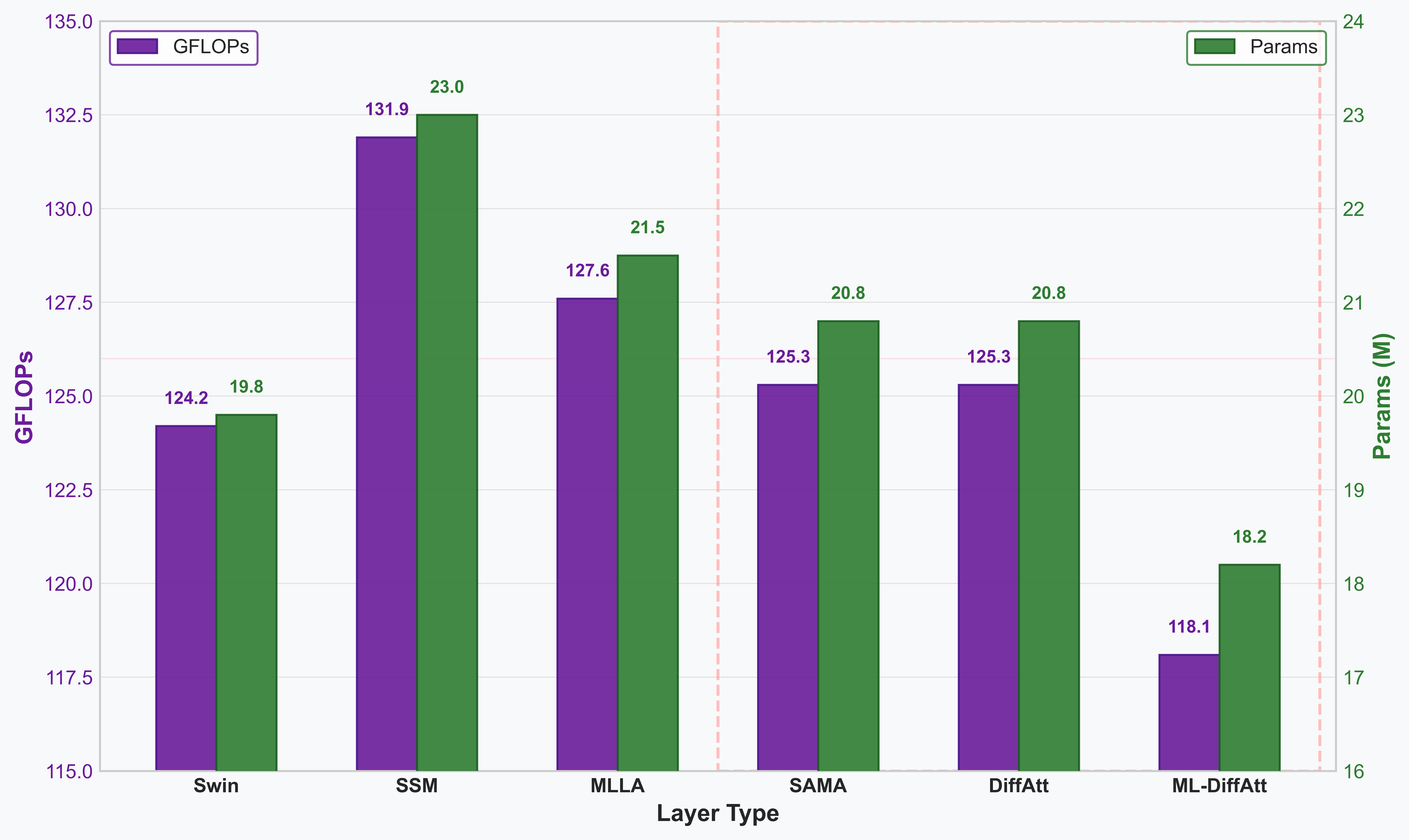}
    \caption{Impact of different types of token mixers on model GFLOPs and parameter count. DiffAtt and ML refer to differential attention and Mamba-like macro structure, respectively.}
    \label{fig5}
\end{figure}

\noindent\textbf{1) Ablation study on SAMA block:} To improve the performance and computational efficiency of Transformer-based token mixer modules, we conducted a comprehensive ablation study on the SAMA module using the BTCV dataset. The detailed designs of the SMM module, MLLA module, and the proposed SAMA module are presented in Figure \ref{fig2}. The experimental results, focusing on segmentation performance and computational complexity, are shown in Figure \ref{fig4} and
Figure \ref{fig5}, with evaluations based on key metrics including DSC, NSD, GFLOPs, and the total number of parameters.

\begin{table}[ht]
\centering
\caption{Ablation studies on the Causal-Resonance Multi-Scale Module, conducted using the BTCV dataset, with DSC (\%) and NSD (\%) as evaluation metrics. The impact of directional multi-view transformations, State Space Modeling (SSM), and causal fusion strategy are independently assessed.}
\label{tab:crmsm_ablation}
\begin{tabular}{lcccccc}
\rowcolor{gray!20}
\textbf{Methods} & \textbf{Multi-View} & \textbf{SSM} & \textbf{Causal Fusion} & \textbf{DSC} & \textbf{NSD} \\
\hline
\rowcolor{gray!10}
SAMA-UNet       & \checkmark & \checkmark & \checkmark &  \textbf{85.38} & \textbf{87.82} \\
(a)             & $\times$   & \checkmark & \checkmark &  85.05 & 87.53 \\
(b)             & \checkmark & $\times$   & \checkmark &  84.93 & 87.48 \\
(c)             & \checkmark & \checkmark & $\times$   &  84.85 & 87.51 \\
\bottomrule
\end{tabular}
\end{table}

As illustrated in Figure \ref{fig4}, our initial modifications to SwinUNETR involved replacing the patch merging, patch expanding, and convolution operations in the encoder stages with Depthwise Separable Convolution modules, while substituting the convolution operations in the skip connections with simple identity mappings. These changes not only improved the DSC by 0.87\% but also significantly reduced the GFLOPs and parameter count from 281.69 and 100.44M to 124.2 and 19.8M, respectively. Building on these modifications, we further replaced the token mixer components with three alternatives: SSM, MLLA, and SAMA modules. 

Replacing the original Shifted Window-based Self-Attention (SWSA) module with SSM and MLLA modules raised the DSC to 82.82\% and 81.81\%, respectively, and improved NSD to 84.23\% and 86.53\%. However, these changes also slightly increased the GFLOPs from 124.2 to 131.9 and 127.6, and the parameters rose from 19.8M to 23.0M and 21.5M, respectively. Using these as baselines, we conducted further analysis of the proposed SAMA module.

For the SAMA module, we first replaced the SWSA module with the Aggregated Attention (Agg) module, which incorporates channel-wise separation and modified local-global pixel-focused attention. This replacement reduced GFLOPs and parameters compared to the SWSA module, while significantly improving the DSC by 5.3\%, as shown in Figure \ref{fig4}. This demonstrates that our proposed attention mechanism achieves linear complexity and effectively extracts both local and global features. On top of this, we incorporated Mamba-like structures into the Agg module, including bypass branches, linear layers, convolution layers, and the SiLU activation function before channel split operation. This addition further boosted the DSC and NSD to 84.37\% and 87.24\%, respectively, outperforming the SMM and MLLA modules, which also employ Mambalike structures. Although this change increased GFLOPs and parameters to 125.3 and 20.8M (as shown in the second to last column of Figure \ref{fig5}), the values remained lower than those of the SMM and MLLA modules. While the linear and convolution layers introduced additional parameters, they also validated the macro-architectural benefits of the Mamba module in enhancing feature extraction capabilities. Finally, we replaced Softmax Attention with Differential Attention. This modification introduced no additional computational cost or parameters but further improved DSC and NSD to 84.88\% and 87.43\%, respectively. These results demonstrate the effectiveness of Differential Attention in suppressing attention noise, assigning higher scores to relevant features while driving irrelevant ones closer to zero. 

Overall, the proposed SAMA module demonstrated superior performance and lower computational complexity compared to SMM and MLLA-based methods when replacing SWSA. This validates the effectiveness of our proposed method.

\noindent\textbf{2) Ablation study on CR-MSM block:} To evaluate the effectiveness of the proposed CR-MSM in capturing long-range, causally aligned spatial dependencies, we conducted a focused set of ablation experiments. These experiments target three core components of CR-MSM: directional multi-view transformations, the State Space Modeling mechanism, and the causal fusion strategy. Each experiment aims to isolate the contribution of one specific design element toward the overall segmentation performance on the BTCV dataset.

The following ablation settings were examined:

    \raisebox{-0.25\height}{\tikz{\node[draw, circle, inner sep=1pt] (a) {a};}} Replacing the directional multi-view transformations with a single original view, to evaluate the impact of spatial orientation diversity on feature representation.
    
   \raisebox{-0.25\height}{\tikz{\node[draw, circle, inner sep=1pt] (b) {b};}} Substituting the SSM block with a standard convolutional layer, thereby removing the structured recurrence modeling designed to capture long-range dependencies.
    
    \raisebox{-0.25\height}{\tikz{\node[draw, circle, inner sep=1pt] (c) {c};}} Replacing the causal fusion mechanism, originally implemented via directional averaging, with simple feature concatenation, eliminating causal alignment during multi-scale integration.

The results, presented in Table III, clearly highlight the importance of each component. In experiment \raisebox{-0.1\height}{\tikz{\node[draw, circle, inner sep=1pt, scale=0.7] {a};}}, removing multi-view transformations resulted in a 0.33\% drop in DSC and 0.29\% drop in NSD, confirming that directional diversity enhances spatial comprehension. In experiment \raisebox{-0.1\height}{\tikz{\node[draw, circle, inner sep=1pt, scale=0.7] {b};}}, replacing SSM with a convolutional layer led to a 0.45\% decrease in DSC and 0.34\% in NSD, emphasizing the advantage of structured recurrence for long-range context modeling. In experiment \raisebox{-0.1\height}{\tikz{\node[draw, circle, inner sep=1pt, scale=0.7] {c};}}, replacing causal fusion with simple concatenation caused a 0.53\% performance loss in DSC and 0.31\% in NSD, underscoring the effectiveness of causal averaging across spatial views.

Overall, these findings validate the architecture of CR-MSM as a robust and efficient module that enhances multi-scale representation learning. Each component, when ablated, results in measurable degradation, confirming that the combined design significantly contributes to segmentation accuracy.

\subsection{Parameters and computational efficiency} 
We also evaluate and compare the computational complexity of each model by measuring the number of trainable parameters and floating-point operations (FLOPs) using the BTCV dataset. As presented in Table 4, our proposed SAMA-UNet achieves a balanced trade-off with 29.03M parameters and 145.67 GFLOPs, maintaining computational efficiency while delivering competitive performance. In contrast, removing the CR-MSM reduces the parameters to 27.2M but increases the computational cost to 356 GFLOPs. Compared to other state-of-the-art models, SAMA-UNet requires significantly fewer parameters and FLOPs than heavy transformer-based models such as LKM-UNet has 189.55M and 993.70G while SwinUNETR has 100.44M and 281.69G. Even against models with encoder-only adaptations like U-Mamba (Enc) and nnUNet, our method remains the most efficient, indicating faster inference potential and reduced memory overhead. These results highlight the computational advantage of SAMA-UNet, making it a suitable choice for real-time or resource-constrained medical imaging applications without compromising segmentation quality.

\begin{table}[ht]
\centering
\caption{\textsc{The FLOPs and Parameters for models on the BTCV dataset. Params denotes parameters.}}
\label{tab:flops_params}
\begin{tabular}{lcc}
\toprule
\rowcolor{gray!9}
\textbf{Methods} & \textbf{Params (M)} & \textbf{GFLOPs} \\
\midrule
nnUNet \cite{34} & 92.48 & 432.13 \\
SwinUNETR \cite{6} & 100.44 & 281.69 \\
LKM-UNet \cite{26} & 189.55 & 993.70 \\
U-Mamba(Enc) \cite{22} & 76.40 & 554.51 \\
\rowcolor{gray!20}
\textbf{SAMA-UNet (ours)} & \textbf{29.03} & \textbf{145.67} \\
\bottomrule
\end{tabular}\label{}
\end{table}

\section{Discussion}\label{sec5}
This study investigates the limitations of State Space Sequence Models in image segmentation tasks, particularly their difficulties in satisfying autoregressive assumptions. Specifically, the absence of causal relationships between image tokens makes it logically infeasible to predict the next token in segmentation tasks based on scanned patch tokens. To address these challenges, we propose the SAMA block and CR-MSM that explore the potential of Mamba-based designs to enhance medical image segmentation from two perspectives. Quantitative and qualitative results across four datasets demonstrate the effectiveness of our model in segmenting multiple organs and surgical instruments. We discuss the technical and clinical impacts of our methods below.

\subsection{Technical impact}
To overcome the limitations of existing State Space Sequence Models (SSMs) in modeling causal relationships between image tokens, we approach the problem from two complementary angles, aligning with the architectural innovations introduced in SAMA-UNet.
To overcome the limitations of existing SSMs in modeling causal relationships between image tokens, we approach the problem from two complementary angles, aligning with the architectural innovations introduced in SAMA-UNet.

In contrast to natural language, image tokens in medical segmentation tasks do not follow an inherently causal sequence. Traditional SSMs, specifically designed for autoregressive modelling, face challenges in effectively capturing spatial dependencies in this context. To address this, we replace SSM-based token mixers in the encoder with our proposed SAMA block. SAMA introduces a hybrid token mixing strategy that utilizes both pixel-focused softmax attention and mamba-inspired macro architectural elements that allow the model to capture both local details and global semantic dependencies without enforcing a causal sequence. Additionally, we integrate differential attention and flash attention mechanisms. The mechanisms inspired by the human foveal vision system allow the network to focus on the most relevant regions while maintaining linear computational complexity. Our experiments, shown in Figures \ref{fig4} and \ref{fig5}, confirm that using the SAMA block instead of SSMs leads to significant performance gains while also reducing the number of model parameters and FLOPs compared to designs based on SSMs.  While attention mechanisms help mix tokens in the encoder, U-shaped segmentation networks also depend a lot on skip connections to share features at different scales. Here, we introduced the CR-MSM to capture implicit causal relationships across resolution levels. In U-shaped networks, encoder features at higher resolutions inherently precede coarser representations and often provide foundational cues for decoder reconstruction. However, conventional multi-scale fusion such as simple concatenation or convolution-based aggregation fails to preserve this directional causality. CR-MSM addresses this concern by applying four directional transformations to the encoder features. A shared SSM block flattens and processes each transformed view, capturing long-range dependencies across spatial orientations. To maintain causal alignment during fusion, we introduce a causal resonance strategy that performs directional averaging rather than naive concatenation. This fusion ensures the model respects the natural order of feature generation.

In the ablation study, Table \ref{tab:crmsm_ablation} confirms the importance of each component. Replacing causal fusion with concatenation results in a decrement in DSC; removing multi-view transformations causes a drop. Together, these results validate our hypothesis that spatial causality is crucial for preserving semantic continuity in medical images.

\subsection{Clinical impact}

From a clinical perspective, the SAMA block and CR-MSM module significantly improve diagnostic precision. By capturing global contextual information, our model accurately identifies structures and boundaries in original images, resulting in precise segmentation maps. These improvements support more accurate diagnoses and facilitate the development of optimized treatment plans. Extensive experiments across various imaging modalities, including MRI, CT, and endoscopy, demonstrate that our method is a robust and versatile tool for medical imaging analysis. The result highlights the potential for integrated multi-modal diagnostics, effectively leveraging complementary information from different modalities to achieve more detailed and precise image analysis. It also offers a scalable solution for leveraging large multi-modal data in medical imaging, aligning with the growing demands for large-scale models and datasets.

Additionally, in radiological workflows, the ability to process and analyze large volumes of high-resolution imaging data quickly can significantly reduce diagnostic time, facilitating faster clinical decision-making. The reduced computational overhead and lower memory footprint of our model make it well-suited for integration into clinical systems with limited hardware resources, further broadening its applicability in real-world medical environments.

\subsection{Limitations and future directions}
Despite the promising results, several challenges remain. First, integrating the Mamba macro-architecture into the SAMA module increases the parameter count, which highlights the need for more effective parameter optimization strategies. Second, extending the model to 3D medical image segmentation requires substantially higher memory, as it triples the number of scanning sequences in the Vision State Space module. To address this, future work can explore memory-efficient 3D processing strategies, such as patch-based or hybrid 2D–3D designs, along with model compression and pruning techniques to reduce resource demands without sacrificing accuracy. Designing lightweight yet expressive modules will be essential to balance efficiency and effectiveness, particularly for real-time 3D clinical applications. In future studies, we aim to refine the architecture, optimize parameter utilization, and investigate memory-efficient methods for scaling to large 3D datasets.

\section{Conclusions}\label{sec6}
In this work, we presented SAMA-UNet, a novel architecture that integrates self-adaptive Mamba-like attention and causal multi-scale fusion for medical image segmentation. The proposed SAMA block and CR-MSM module enable the network to capture both fine local details and broad global context with improved efficiency. Extensive experiments on four benchmark datasets (BTCV, ACDC, EndoVis17, and ATLAS23) demonstrate that SAMA-UNet consistently outperforms CNN, Transformer, and Mamba-based models across CT, MRI, and endoscopic imaging. Future work will focus on developing lighter variants, exploring memory-efficient 3D adaptations, and enabling real-time deployment in clinical workflows.


\vspace{10pt} 
\noindent\textbf{Conflict of interest:} The authors declare no competing interests.

\noindent\textbf{Data availability:} The dataset is publicly available in the cited references: BTCV \cite{29}, ACDC \cite{30}, EndoVis17 \cite{31}, and ATLAS23 \cite{32}. More details can be shared upon request.

\noindent\textbf{Code availability:} The code is available at \href{https://github.com/sqbqamar/SAMA-UNet}{https://github.com/sqbqamar/SAMA-UNet}.

\vspace{6pt}

\backmatter

\bmhead{Acknowledgements}
This work was supported and funded by the Deanship of Scientific Research at Imam Mohammad Ibn Saud Islamic University (IMSIU) (grant number IMSIU-DDRSP2501)

\bmhead{Funding Statement}
This work was supported and funded by the Deanship of Scientific Research at Imam Mohammad Ibn Saud Islamic University (IMSIU) (grant number IMSIU-DDRSP2501).

\bibliography{SAMA}

\end{document}